\documentclass[prl,twocolumn,showpacs,superscriptaddress,preprintnumbers,amsmath,amssymb,groupedaddress]{revtex4}

\usepackage{graphicx}
\usepackage{dcolumn}
\usepackage{bm}
\usepackage{color}

\begin{document}

\preprint{}

\title{Magnetic field-induced phase transitions in a weakly coupled s = 1/2 quantum spin dimer system Ba$_{3}$Cr$_{2}$O$_{8}$
}

\author{M. Kofu}
\affiliation{
Department of Physics, University of Virginia, Charlottesville, Virginia 22904, USA
}%

\author{H. Ueda}
\affiliation{
Institute for Solid State Physics, University of Tokyo, Kashiwa 277-8581, Japan
}%

\author{H. Nojiri}
\author{Y. Oshima}
\author{T. Zenmoto}
\affiliation{
Institute for Materials Research, Tohoku University, Sendai, Miyagi 980-0821, Japan
}%

\author{K. Rule}
\author{S. Gerischer}
\affiliation{
Helmholtz-Zentrum Berlin, Berlin 14109, Germany.
}%

\author{B. Lake}
\affiliation{
Helmholtz-Zentrum Berlin, Berlin 14109, Germany.
}%
\affiliation{
Institut f\"{u}r Festk\"{o}rperphysik, Technische Universit\"{a}t Berlin, Hardenbergstra\ss e 36, 10623 Berlin, Germany.
}%

\author{C. D. Batista}
\affiliation{
Theoretical Division, Los Alamos National Laboratory, Los Alamos, New Mexico, 87545 USA
}%

\author{Y. Ueda}
\affiliation{
Institute for Solid State Physics, University of Tokyo, Kashiwa 277-8581, Japan
}%

\author{S.-H. Lee}
\email{shlee@virginia.edu}
\affiliation{
Department of Physics, University of Virginia, Charlottesville, Virginia 22904, USA
}%

\date{\today}

\begin{abstract}
By using bulk magnetization, electron spin resonance (ESR), heat capacity, and neutron scattering techniques, we characterize the thermodynamic and quantum phase diagrams of Ba$_3$Cr$_2$O$_8$. Our ESR measurements indicate that the low field paramagnetic ground state is a mixed state of the singlet and the S$_z$ = 0 triplet for $H \perp c$. This suggests the presence of an intra-dimer Dzyaloshinsky-Moriya (DM) interaction with a DM vector perpendicular to the c-axis. 
\end{abstract}

\maketitle


Over the last two decades, several exotic collective phenomena have been found in quantum ($s = 1/2$) magnets~\cite{Sachdev08}. One of the well-known examples is the condensation of magnons induced by application of an external magnetic field~\cite{Sachdev08,Giamarchi08}. The field-induced quantum phase transition has been experimentally observed in coupled quantum dimer systems, such as TlCuCl$_3$~\cite{Nikuni_2000, TlCuCl3_n} and BaCuSi$_2$O$_6$~\cite{Sebastian06}. If the field is applied along an axis of U(1) 
symmetry, the induced quantum critical point (QCP)  belongs to the Bose-Einstein condensation (BEC) universality class~\cite{Giamarchi99}. 
One signature of the BEC universality class is the power law for the increase in the  transition temperature, $T_c$ close enough to the 
critical field $H_{c1}$ : $T_c \propto [ H - H_{c1} ]^{1/\phi}$, with the critical exponent 
$\phi = d/2$ for spatial dimensionality $d \geq 2$~\cite{Giamarchi99}. The BEC value of $\phi$ has been found in the two-dimensionally (2d) coupled dimer system BaCuSi$_2$O$_6$~\cite{Sebastian06}, a frustrated 2d antiferromagnet Cs$_2$CuCl$_4$~\cite{Radu05}, and a three-dimensionally coupled s = 1 chain NiCl$_2$-4SC(NH$_2$)$_2$~\cite{Zapf06}. However, values of $\phi$ higher than 3/2 have been reported for the three-dimensionally coupled quantum dimer compounds such as TlCuCl$_3$ ($\phi = 2.2(1)$) and KCuCl$_3$ ($\phi = 2.3(1)$) for 1 K $< T <$ 5 K.
Quantum Monte Carlo (QMC) simulations of U(1) invariant models~\cite{Wessel01,Kawashima04} have shown that 
$\phi$ can be easily overestimated by fitting the non-universal regime. 
On the other hand, when a narrow enough temperature range $\Delta T$ is considered ($\Delta T \simeq 0.4 T_{max}$ for an s = 1/2 XXZ magnet), 
the QMC results yield the expected $\phi = d/2$ exponent~\cite{Kawashima04}. $\Delta T$ is the typical size of the 
quantum critical region.

A question that arises is how robust is the BEC quantum critical behavior against the presence of magnetic anisotropy. A case in point here is the three-dimensionally coupled quantum dimer system Ba$_3$Cr$_2$O$_4$ where the tetrahedrally co-ordinated Cr$^{5+} (3d^1; s = 1/2)$ form dimers along the crystallographic $c$-axis which in turn are coupled into a triangular lattice in the $ab$-plane. Unlike the related material, Ba$_3$Mn$_2$O$_8$ that has the orbitally nondegenerate Mn$^{5+} (3d^2; s = 1)$ ions, Ba$_3$Cr$_2$O$_4$ undergoes a structural phase transition at 70 K due to the $e_g$ Jahn-Teller active Cr$^{5+}$ ion that leads to spatially anisotropic three-dimensional exchange interactions and lifts the frustration of the triangular lattice~\cite{Kofu09}. Without an external magnetic field, Ba$_3$Cr$_2$O$_4$ does not exhibit any long range order down to 1.5~K.  When an external magnetic field is applied, however, the bulk magnetization data shows a field-induced transition into a magnetically ordered state at $H_{c1} \simeq 12$ T and another transition into a fully polarized state at $H_{c2} \sim$ 23 T~\cite{Nakajima_2007}.

In this letter, we report our systematic studies of the field-induced phase transitions in Ba$_3$Cr$_2$O$_8$, using bulk magnetization, specific heat, electron spin resonance (ESR), and both elastic and inelastic neutron scattering techniques. 
Our ESR data showed that Ba$_3$Cr$_2$O$_8$ realizes a system of three-dimensionally coupled quantum dimers with weak inplane Dzyaloshinsky-Moriya (DM) interactions less than 0.1 meV. 
Our specific heat and elastic neutron scattering data show that the $H-T$ phase boundary of the field-induced canted 
antiferromagnetic (AFM) state near the lower critical field follows a power law $T_c \propto \mid H_{c} (T) - H_{c} (0) \mid^{1/\phi}$ 
that is consistent with $\phi = d/2$ for 30mK $< T <$ 1 K. 

\begin{figure}[tb]
\begin{center}
\includegraphics[width=\hsize]{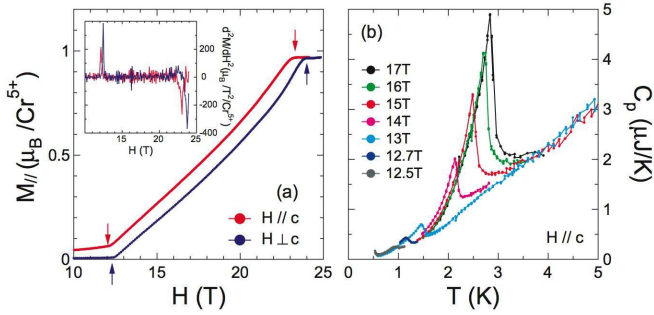}
\caption{(color online) (a)~Bulk magnetization~($M$) as a function of an external pulsed magnetic field $(H)$ applied parallel and perpendicular to the $c$-axis, taken at $T=0.4$~K. Arrows indicate upper and lower critical fields determined by the positions of the peaks that appear in the second derivative, $d^{2}M/dH^{2}$, as a function of $H$ as shown in the inset. (d)~Temperature dependence of the specific heat under various constant steady magnetic fields along the $c$-axis, from $H =$ 12.5 T to 17 T. The data are consistent with Aczel {\it et al.}'s recent results reported in Ref. \cite{Aczel09}.}
\label{fig:M_Cp}
\end{center}
\end{figure}

Single crystals of Ba$_3$Cr$_2$O$_8$ were grown at the Institute for Solid State Physics~(ISSP), University of Tokyo as well as at the University of Virginia, using floating zone image furnaces. Bulk magnetization, electron spin resonance and specific heat measurements were performed at the Institute for Materials Research~(IMR) of Tohoku University using small single crystals, a 15 mg piece for $H \perp c$ and two co-mounted crystals of total weight of 20 mg for $H \parallel c$. Elastic and inelastic neutron scattering measurements were carried out at the cold-neutron triple-axis spectrometer V2 at the Helmholtz-Zentrum Berlin~(HZB). A larger single crystal of 2.5 g was used for the measurements and was oriented with $(h,h,l)$ as the scattering plane. A 15~T magnet with dilution insert was used. A flat analyzer configuration with a final neutron energy of $E_f = $3.7 and 5~meV and a horizontally focusing configuration with $E_f = 5$ meV were used for elastic and inelastic neutron scattering measurements, respectively. A cooled Be filter was placed before the analyzer to eliminate higher order neutron contaminations. Horizontal collimations were guide-60'-open-open at V2.


Figure~\ref{fig:M_Cp}(a) shows bulk magnetization~($M_{\parallel }$) measured using a pulsed magnet at $T=0.4$~K and for fields up to 25~T. Upon ramping up, the ferromagnetic longitudinal component of the magnetic moment, $M_{\parallel }$, increases rapidly at $H_{\rm c1}$ = 12.15(5)~T when $H\perp c$, as previously reported, and at $H_{\rm c1}$ = 12.45(5)~T when $H \parallel c$. Upon further increasing $H$, $M_{\parallel }$ increases linearly up to a saturated value, $M_s = g s \mu_B/$Cr$^{5+}$ at $H_{c2}= 23.05(5)$~T for $H\perp c$ and at $H_{c2} = 23.70(5)$ for $H \parallel c$. The critical fields marked by the arrows are determined from the second derivative, $d^2M/dH^2$. Fig.~\ref{fig:M_Cp}(b) shows specific heat data as a function of temperature measured at various fields up to 17 T applied along the $c$-axis.  
At 12.5 T just above $H_{\rm c1}$, a weak $\lambda$-type anomaly is observed at 0.94~K. As $H$ increases up to 17 T, the anomaly becomes more prominent and the transition temperature also increases up to 2.84~K.

\begin{figure}[tb]
\begin{center}
\includegraphics[width=\hsize]{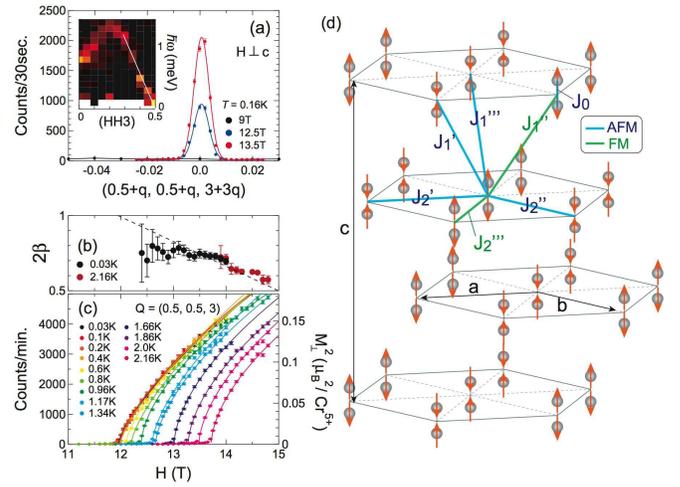}
\caption{(color online) (a)~Elastic neutron scattering data around the magnetic (0.5,0.5,3) reflection, measured at 0.15 K and with three different magnetic fields, $H =$ 9 T (black), 12.5 T (blue), and 13.5 T (red circles). The inset shows the color contour map of the inelastic neutron scattering intensity of the $S_z = 1$ triplet excitation measured with $H = 14.8$ T. (b) The critical exponent 2$\beta$ obtained from the data shown in (c). See the text for detailed description. (c)~$H$-dependence of the magnetic (0.5,0.5,3) Bragg reflection measured at various temperatures from 30 mK to 2.16 K. $H$ was perpendicular to the $c$-axis. The solid lines are fits to a power-law.
(c)~Schematic figure of the structure of the transverse moments in the canted AFM phase of Ba$_3$Cr$_2$O$_8$ for $H\perp c$, obtained from our magnetic Bragg scattering data. Arrows represent the transverse magnetic moments along the $c$-axis. The magnetic unit cell is $2a\times 2b\times c$.}
\label{fig:neutron}
\end{center}
\end{figure}

In order to investigate the nature of the field-induced state above $H_{\rm c1}$, we performed elastic neturon scattering measurements as a function of magnetic field applied perpendicular to the $c$ axis and temperature.
As shown in Fig.~\ref{fig:neutron}(a), for $H \geq H_{\rm c1}$ magnetic Bragg peaks appear at $Q=(0.5, 0.5, l)$ with $l = 3n$ (integer $n$).
This, combined with the bulk magnetization data, clearly indicates that the field-induced phase is a canted antiferromagnetic state.
As $H$ increases, the magnetic Bragg intensity increases. The $H$-dependence of the intensity has been measured at various $T$s, as shown in Fig.~\ref{fig:neutron}(c). Upon increasing $T$ from 30 mK to 2.1 K, $H_{\rm c1}$ increases from 11.95(2)~T to 13.72(1)~T. The $H$-dependent intensity is proportional to the square of the antiferromagnetic transverse component of the magnetic moment, $M_{\perp}^2$ and follows the power law, $M_{\perp}^2 \propto (H - H_{\rm c1})^{2\beta}$. The solid lines are the fits to the power law using all data available, yielding $2\beta = 0.73(3)$ for 30 mK and 0.57(3) for 2.1 K. 
The experimental data is consistent with the expected crossover between a lower classical value of $\beta$ at high temperature ($\beta \simeq 0.348$ for a 3d-XY universality class) and the higher mean field value  $\beta = 1/2$ expected for the  QPT ($T=0$). Note that the effective dimensionality $D=d+z \geq 4$ for the spatial dimensionality $d=3$ and the dynamical exponent $z \geq 1$ ($z=2$ for a BEC-QCP).


The magnetic Bragg scattering was measured at three different reflections $Q=(0.5, 0.5, l)$ with $l = 0, 3, 6$ at $T = 0.16$ K and $H = 13.5$~T. As shown in Table I, their relative intensities could be reproduced best by a magnetic structure where the transverse components order along the $c$-axis in the collinear structure that is illustrated in Fig.~\ref{fig:neutron}(d). Note that $M_{\perp}$ is ordered ferromagnetically along the [110] direction and antiferromagnetically along the [100]/[010] directions, which is consistent with the spatially anisotropic exchange interactions previously determined by the dispersion relations of the triplet excitations with $H=0$~T~\cite{Kofu09}. As shown in the upper left inset of Fig.~\ref{fig:neutron}(a), for $H > H_{\rm c1}$ the triple excitation with $S_z = 1$ condenses and becomes a gapless Goldstone mode with a spin wave velocity of 2.61(14) meV/\AA$^{-1}$. No anisotropy gap could not be detected within the experimental energy resolution of $\Delta E=0.15$ meV, setting the upper limit of the gap size to 0.07 meV.

\begin{table}[tb]
\caption{Experimental magnetic Bragg intensities measured at $T=0.16$~K and $H=13.5$~T and corrected for the Lorentz factor, ($\sin{2\theta}$). The calculated intensities for two different magnetic structures, one with $\bm M_\perp \parallel $[001] and the other with $\bm M_\perp \parallel $[110] are also listed. }
\begin{ruledtabular}
\begin{tabular}{cccc}
($h, k, l$) & $I_{\rm obs}$ & $I_{\rm calc}$ ($\bm M_\perp \parallel $[001]) & $I_{\rm calc}$ ($\bm M_\perp \parallel$ [110])\\
\hline
(0.5, 0.5, 0) & 0$\pm$2.8 & 0 & 0\\ 
(0.5, 0.5, 3) & 1646.8$\pm$32.2 & 1646.8 & 1646.8\\ 
(0.5, 0.5, 6) & 55.5$\pm$3.6 & 81.8 & 327.0\\ 
\end{tabular}%
\label{tab:table}
\end{ruledtabular}
\end{table}%

\begin{figure}[tb]
\begin{center}
\includegraphics[width=\hsize]{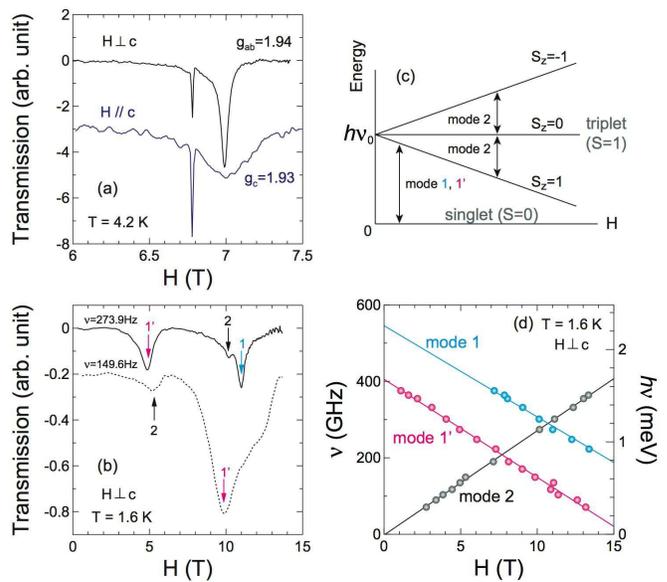}
\caption{(color online) Electron spin resonance transmission spectra as a function of external magnetic field (a) at 4.2 K with $H \perp c$ (blue line) and $H \parallel c$ (red line), obtained with the frequency of $\nu = 190$~GHz, and (b) at 1.6 K with $H \perp c$ and with two different frequancies $\nu = 149.6$~GHz (dotted line) and 273.9 GHz (solid line). (c) Energy level scheme of the nominal singlet and triplet states as a function of $H$. (d) Frequency dependence of the peak positions of the three modes, 1, 1$^{\prime}$, 2, in $H$, which are described in the text. 
}
\label{fig:ESR}
\end{center}
\end{figure}

In order to find out what kind of perturbations may exist in the spin Hamiltonian of this system, we have performed electron spin resonance (ESR) measurements as a function of $H$ and the frequency. These measurements probe $\Delta S = 0$ and $\Delta S_z = 1$ transition processes~\cite{Nojiri06}, such as between the spin-triplet states when they are split by the external magnetic field. Firstly, in order to estimate the gyromagnetic ratio, $g$, for Ba$_3$Cr$_2$O$_8$, the ESR measurements were done with a standard DPPH (diphenyl-picri-hydrazyl) sample that has the $g$-factor of 2.00. The ESR transmission spectra are shown as solid lines in Fig.~\ref{fig:ESR}(a). 
The sharp peak at $H =6.8$~T comes from the standard DPPH while the broad peaks at $H \sim 7$~T come from Ba$_3$Cr$_2$O$_8$. The fact that the peak positions of the broad peaks appear at  $H$  close to but higher than the peak position of DPPH, $H_{\rm DPPH}$, indicates that the $g$-factor of Ba$_3$Cr$_2$O$_8$ is slightly lower than 2: $g_{ab} = 2 H_{\rm DPPH} / H_{ab} =$ 1.94 and $g_c = 2 H_{\rm DPPH} / H_{c} =$ 1.93. The order of magnitude of the symmetric exchange anisotropy can be estimated as $J_0 (\Delta g/g)^2  \simeq 1 \mu$eV.

More ESR measurements with $H \perp c$ were done without DPPH using various frequencies some of which are shown in
Fig.~\ref{fig:ESR}(b). Three ESR modes, labeled by 1, 1$^{\prime}$, and 2, were identified. The mode 2 is due to the expected $\Delta S = 0$ transitions among the triplet states split under $H$ as shown in Fig.~\ref{fig:ESR}(c), which energy increases linearly with $H$ (the black symbols in Fig.~\ref{fig:ESR}(d)). On the other hand, for the other two modes, 1 and 1$^{\prime}$, their characteristic energies decrease with increasing $H$, following $h\nu = h\nu_0 - g\mu_B H$ with $\nu_0 = 546$(9)~GHz $= 2.2$ meV (mode 1) and 406(4)~GHz $= 1.62$ meV (mode 1$^{\prime}$). The $\nu_0$s correspond to the energies of the singlet-to-triplet excitations that are 2.22(16) meV and 1.76(17) meV, respectively, determined by $\hbar\omega (\bm Q) \simeq  \sqrt{J_0^2+J_0~\gamma(\bm Q)}$ where $\gamma(\bm Q=0)=2(J_1^{\prime}+J_1^{\prime \prime}+J_1^{\prime \prime \prime})\cos(\frac{2}{3}n\pi) + 2(J_2^{\prime}+J_2^{\prime \prime}+J_2^{\prime \prime \prime})+2(J_4^{\prime}+J_4^{\prime \prime}+J_4^{\prime \prime \prime})\cos(\frac{2}{3}n\pi)$ with $n=0$ and $\pm 1$ for the acoustic and optical modes, respectively, and $J$s are determined by the previous neutron study. These indicate that the modes 1 and $1^{\prime}$ are due to the transition from the ground state to the $\mid S, S_z> = \mid 1,1>$ state. Thus, the ground state of Ba$_3$Cr$_2$O$_8$ must be a mixed state of the singlet and the $S_z=0$ triplet state, $\mid \Psi>_{GS} = \mid 0,0> +f \mid 1,0>$ where $f$ is a mixing coefficient. Such a mixing is possible if there is a Dzyaloshinsky-Moriya (DM) interaction term, $\vec{D} \cdot (\vec{s}_i \times \vec{s}_j)$ with $\vec{D} \parallel \vec{H} \perp c$. When $H \parallel c$, the DM term, $\vec{D} \perp c$ would mix $\mid 0,0>$ with $\mid 1,\pm 1>$, which might be the reason for the weak non-zero bulk magnetization for $H < H_{c1}$ when $H \parallel c$ as shown in Fig. 1 (a). The value of 
$M (H \sim H_{c1})/M (H = H_{c2}) \leq 0.05$ should be order of $\mid f\mid^2 \sim \mid D/h\nu (H_{c1})\mid^2$ where $h\nu (H_{c1}) \sim 0.4$ meV is the energy of the ESR signal at $H_{c1}$, yielding $\mid \vec{D} \mid \leq 0.1$ meV. The existence of the DM interactions suggests that the inversion symmetry is broken at the center of the dimer and the crystal symmetry is lower than the previously reported C2/c symmetry~\cite{Kofu09, Chapon08}.

\begin{figure}[t]
\begin{center}
\includegraphics[width=\hsize]{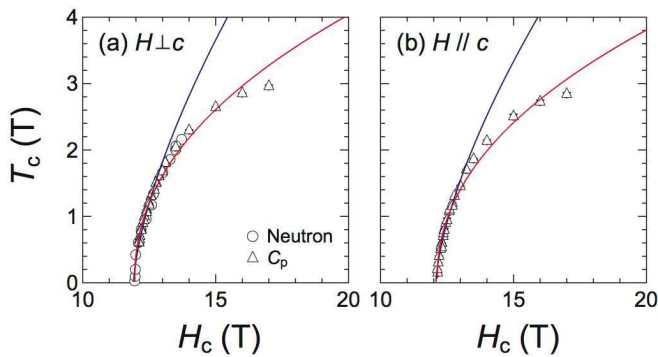}
\caption{(color online) $H-T$ phase diagram (a)~for $H \perp c$ determined from the elastic neutron scattering data (circles) shown in Fig. 2 (b) and specific heat data (triangles), 
and (b)~for $H \parallel c$ determined from the heat capacity data shown in Fig. 1 (b). The lines are fits to a power law (see the text for details).}
\label{fig:phase_diagram}
\end{center}
\end{figure}

The $H-T$ phase diagram was constructed from the elastic neutron and specific heat data for $H\perp c$ (Fig.~\ref{fig:phase_diagram} (a)) and the specific heat data for $H\parallel c$ (Fig.~\ref{fig:phase_diagram} (b)). 
Note that there is only one antiferromagnetic phase in both cases. This contrasts to the related compound Ba$_3$Mn$_2$O$_8$ where, for $H\perp c$, a second antiferromangetic phase exists over a narrow region of field, which was accounted for by the existence of a single-ion anisotropy~\cite{Samulon08}. 
The phase boundary near $H_{\rm c1}$ was fitted to a power law, $T_c \propto \mid H_{c} (T) - H_{c} (0) \mid^{1/\phi}$. The optimal fits were obtained with $H_{c1} (0) = 11.95(1)$ T and $\phi = 2.22(5)$ for $H \perp c$, and with $H_{c1} (0) = 12.13(1)$~T and $\phi = 2.21(6)$ for $H \parallel c$ (red lines in Fig.~\ref{fig:phase_diagram}). On the other hand, when $\phi$ is fixed to the BEC value of $d/2 = 1.5$, the phase boundaries were equally well reproduced up to $\sim$ 1~K with $H_{c1} (0) = 11.92(1)$ T for $H \perp c$ and $H_{c1} (0) = 12.08(1)$ T for $H \parallel c$, as shown by the blue lines. A previous Monte Carlo simulation study has shown that the BEC regime extends up to $\sim$ 0.4 $T_{max}$ where $T_{max}$ is the maximum temperature of the canted AFM regime~\cite{Kawashima04}. This is consistent with our results since $T_{max} \sim 3$ K.

The mixing of the singlet and triplet states in the ground state has also been found in other couple dimer systems such as KCuCl$_3$~\cite{Tanaka98} and SrCu$_2$(BO$_3$)$_2$~\cite{Nojiri03} and were ascribed for the existence of DM interactions. 
The DM interaction breaks the U(1) symmetry for $H \parallel c$ and consequently it is a relevant perturbation for the BEC-QCP. The Ising-like QCP expected for this field orientation has an exponent $\phi=2$ that should appear below a characteristic temperature $T_0$ that depends on the magnitude of the DM anisotropy. If If $T_0 \ll 0.4 T_{max}$  one should still observe the BEC exponent in the window $T_0 \lesssim T \lesssim 0.4T_{max}$~\cite{Mayafre00}. Our analysis of the measured $T_c (H)$ curve shows that this data alone is not enough to distinguish between the two universality classes. 


In summary, using several different techniques such as ESR, specific heat and neutron scattering, we have characterized in detail the magnetic interactions and constructed the $H-T$ phase diagram of the newly discovered quantum dimer system Ba$_3$Cr$_2$O$_8$. Our results show that near the lower critical field, $H_{c1}$, the $H-T$ phase diagram between the disordered and canted AFM states can be reproduced by a power law $T_c \propto \mid H_{c} (T) - H_{c} (0) \mid^{1/\phi}$ that is consistent with $\phi = d/2$ up to 1 K $\sim 0.4~T_{max}$. Further measurements on other physical quantities near the QCP may be needed to single out the universality class that dominates in this temperature range.


We thank M. Tachiki, S. Haas, Y. B. Kim, S. Ishihara,  O. Nohadni for helpful discussions, C. Stock and V. G. Sakai for crystal alignment for neutron scattering measurements. Work at the University of Virginia was supported by the U.S. DoE through DE-FG02-07ER46384.

\end{document}